\documentclass[article,preprint,groupedaddress]{revtex4}
\usepackage{epsfig}
\usepackage{graphicx}
\usepackage{amssymb}

\begin{document}
\title{No-go theorem for static boson stars}
\author{Shahar Hod}
\affiliation{The Ruppin Academic Center, Emeq Hefer 40250, Israel}
\affiliation{ }
\affiliation{The Hadassah Institute, Jerusalem 91010, Israel}
\date{\today}

\begin{abstract}
\ \ \ It is proved that self-gravitating static scalar fields whose
self-interaction potential $V(\psi^2)$ is a monotonically increasing
function of its argument cannot form spherically symmetric
asymptotically flat bound matter configurations. Our compact theorem
rules out, in particular, the existence of spatially regular static
boson stars made of nonlinear massive scalar fields.
\end{abstract}
\bigskip
\maketitle

\section{Introduction}

The physical and mathematical properties of the composed
Einstein-scalar system have attracted much attention during the last
five decades. Intriguingly, while the elegant no-hair theorems of
Chase \cite{Cha}, Bekenstein \cite{Bek1}, and Teitelboim \cite{Teit}
(see also \cite{Heu,Bek2,BekMay,Bek20,Hodnmc} and references
therein) have revealed the important fact that {\it static} scalar
fields cannot form hairy black-hole configurations with regular
event horizons \cite{NoteV}, it has recently been proved that {\it
stationary} spatially regular massive scalar hair can be supported
in asymptotically flat spinning black-hole spacetimes
\cite{Hodrc,Herkr}.

Boson stars, horizonless self-gravitating massive scalar
configurations, are also known to exist as {\it stationary}
solutions of the non-linearly coupled Einstein-scalar field
equations (see \cite{Fra,Lie} for excellent reviews). These compact
objects have been considered in the physics literature as exotic
horizonless black-hole mimickers, as possible sources of dark matter
configurations, and as self-gravitating compact objects in binary
systems \cite{Fra,Lie}.

One naturally wonders whether horizonless compact boson stars can be
constructed from {\it static} (rather than stationary
\cite{Fra,Lie}) self-interacting scalar fields? Derrick's well-known
theorem \cite{Der} guarantees that, in $(3+1)$-dimensional flat
spacetimes, spatially regular static scalar field configurations
with positive definite energy density cannot exist. The main goal of
the present paper is to extend this no-go theorem for static boson
stars to the regime of self-gravitating scalar field configurations
in curved spacetimes.

To this end, in the present paper we shall explore the physical and
mathematical properties of the static sector of the nonlinearly
coupled Einstein-scalar field equations. In particular, below we
shall explicitly prove that, within the framework of classical
general relativity, self-gravitating static scalar fields with
generic self-interaction potentials {\it cannot} form spatially
regular horizonless matter configurations (boson stars)
\cite{NoteTej,Tej}.

\section{The no-go theorem for spherically symmetric asymptotically flat static boson stars}

We study the physical and mathematical properties of spherically
symmetric static matter configurations made of a self-gravitating
real scalar field $\psi$ whose action is given by
\cite{BekMay,Noteunit,NoteSEH}
\begin{equation}\label{Eq1}
S=S_{EH}-{1\over2}\int\Big[\partial_{\alpha}\psi\partial^{\alpha}\psi+V(\psi^2)\Big]\sqrt{-g}d^4x\
,
\end{equation}
where the nonlinear scalar self-interaction potential $V(\psi^2)$ is
assumed to be a monotonically increasing and a positive semidefinite
function of its argument:
\begin{equation}\label{Eq2}
V(0)=0\ \ \ \ \text{with}\ \ \ \ \dot V\equiv
{{d[V(\psi^2)]}\over{d(\psi^2)}}\geq0\ .
\end{equation}
Note, in particular, that the physically interesting case of an
asymptotically flat self-gravitating massive scalar field
configuration with $\dot V=\mu^2\geq0$ is covered by this functional
form of the nonlinear self-interaction potential.

The spherically symmetric spacetime describing the self-gravitating
static scalar field configuration is characterized by the curved
line element \cite{BekMay,Notesch}
\begin{equation}\label{Eq3}
ds^2=-e^{\nu}dt^2+e^{\lambda}dr^2+r^2(d\theta^2+\sin^2\theta
d\phi^2)\ ,
\end{equation}
where the metric functions $\{\nu,\lambda\}$ depend on the areal
coordinate $r$. Spatially regular spacetimes are characterized by
the near-origin behavior \cite{Hodt1}
\begin{equation}\label{Eq4}
e^{\lambda}(r\to 0)=1+O(r^2)\ \ \ \ ; \ \ \ \ 0<e^{\nu}<\infty\
\end{equation}
with \cite{Notetag,Notenbo}
\begin{equation}\label{Eq5}
\lambda'(r\to0)\to0\ \ \ \ ; \ \ \ \ \nu'(r\to0)\to0\  .
\end{equation}
In addition, asymptotically flat spacetimes describing spatially
regular matter configurations of finite mass are characterized by
the simple dimensionless functional relations \cite{BekMay}
\begin{equation}\label{Eq6}
\nu(r\to\infty)=O(M/r)\ \ \ \ ; \ \ \ \ \lambda(r\to\infty)=O(M/r)\
\end{equation}
at spatial infinity, where $M$ is the total ADM mass (as measured by
asymptotic observers) of the spatially regular field configuration.

The action (\ref{Eq1}) of the self-interacting scalar field yields
the characteristic nonlinear Klein-Gordon wave equation
\cite{BekMay}
\begin{equation}\label{Eq7}
\partial_{\alpha}\partial^{\alpha}\psi-\dot V\psi=0\  ,
\end{equation}
which, taking cognizance of the curved line element (\ref{Eq3}) that
characterizes the spherically symmetric static matter configuration,
can be written in the form \cite{BekMay}
\begin{equation}\label{Eq8}
\psi{''}+{1\over2}\big({{4}\over{r}}+\nu{'}-\lambda{'}\big)\psi{'}-e^{\lambda}\dot
V\psi=0\  .
\end{equation}

The action (\ref{Eq1}) also yields the functional expression
\cite{BekMay}
\begin{equation}\label{Eq9}
\rho=-T^{t}_{t}={1\over2}\big[e^{-\lambda}(\psi{'})^2+V(\psi^2)\big]\
\end{equation}
for the energy density of the self-interacting scalar field. Taking
cognizance of Eqs. (\ref{Eq2}) and (\ref{Eq9}), and using the fact
that physically acceptable spacetimes are characterized by finite
energy densities \cite{BekMay}, one deduces that the gradient of the
scalar eigenfunction is finite \cite{Noteeg}:
\begin{equation}\label{Eq10}
\psi'(r)<\infty\  .
\end{equation}
In addition, taking cognizance of Eqs. (\ref{Eq2}), (\ref{Eq6}), and
(\ref{Eq9}), and using the fact that finite mass matter
configurations in asymptotically flat spacetimes are characterized
by the simple asymptotic behavior \cite{Hodasm}
\begin{equation}\label{Eq11}
r^3\rho(r)\to0\ \ \ \ \text{for}\ \ \ \ r\to\infty\  ,
\end{equation}
one deduces that the self-gravitating bound-state scalar
configurations are characterized by the asymptotic functional
behavior
\begin{equation}\label{Eq12}
\psi(r\to\infty)\to0\  .
\end{equation}

We shall now prove that the eigenfunction $\psi(r)$, which
characterizes the spatial behavior of the spherically symmetric
self-gravitating static scalar fields, cannot be a monotonic
function of the radial coordinate $r$. Taking cognizance of Eqs.
(\ref{Eq4}), (\ref{Eq5}), and (\ref{Eq8}), one finds the near-origin
radial scalar equation
\begin{equation}\label{Eq13}
\psi{''}+{{2}\over{r}}\psi{'}-\dot V\psi=0\ .
\end{equation}
The physically acceptable solution of (\ref{Eq13}) which respects
the relation (\ref{Eq10}) is characterized by the small-$r$
(near-origin) functional behavior
\begin{equation}\label{Eq14}
\psi(r\to0)=a\big[1+{1\over6}\dot V(a^2)\cdot r^2\big]+O(r^3)\  ,
\end{equation}
where $a$ is a constant. Using Eqs. (\ref{Eq2}) and (\ref{Eq14}),
one obtains the near-origin functional relations
\cite{Noteano,Notevdt}
\begin{equation}\label{Eq15}
\psi\psi'(r\to0)=0\ \ \ \ ; \ \ \ \ \psi\psi''(r\to0)>0\ .
\end{equation}

From the small-$r$ and large-$r$ functional behaviors (\ref{Eq12})
and (\ref{Eq15}) of the radial scalar eigenfunction $\psi(r)$, which
characterizes the spatially regular static matter configurations,
one deduces that $\psi(r)$ must have (at least) one extremum point,
$r=r_{\text{peak}}$, with the characteristic functional relations
\begin{equation}\label{Eq16}
\{\psi\neq0\ \ \ ; \ \ \ \psi{'}=0\ \ \ ; \ \ \
\psi\cdot\psi{''}<0\}\ \ \ \ \text{for}\ \ \ \ r=r_{\text{peak}}\ .
\end{equation}
In particular, at the extremum point (\ref{Eq16}), the radial scalar
equation (\ref{Eq8}) yields the remarkably simple relation
\begin{equation}\label{Eq17}
\psi\psi{''}=e^{\lambda}\dot V\psi^2\ \ \ \ \text{for}\ \ \ \
r=r_{\text{peak}}\  .
\end{equation}
Taking cognizance of Eqs. (\ref{Eq2}) and (\ref{Eq16}), one learns
that the functional expression on the l.h.s of (\ref{Eq17}) is
negative whereas the functional expression on the r.h.s of
(\ref{Eq17}) is non-negative. Thus, the characteristic differential
relation (\ref{Eq8}) for the self-gravitating static matter
configurations {\it cannot} be respected at the extremum point
(\ref{Eq16}) of the scalar eigenfunction.

We therefore conclude that spherically symmetric asymptotically flat
bound-state matter configurations (boson stars) cannot be
constructed from spatially regular static scalar fields whose
self-interaction potential $V(\psi^2)$ is a monotonically increasing
function of its argument.

\section{Summary}

In this compact analysis of the static sector of the nonlinearly
coupled Einstein-scalar field equations, we have explicitly proved
that {\it if} a spherically symmetric spatially regular boson star
can be constructed from a {\it static} (rather than a stationary)
scalar field, then the corresponding radial scalar eigenfunction
$\psi(r)$ cannot be a monotonic function of the areal coordinate
$r$. In particular, at its extremum point (\ref{Eq16}), the scalar
field eigenfunction is characterized by the simple functional
relation (\ref{Eq17}). However, one immediately realizes that this
relation is in {\it contradiction} with the functional identity
(\ref{Eq8}) [see Eqs. (\ref{Eq2}) and (\ref{Eq16})] which
characterizes the radial behavior of the nonlinear self-gravitating
static scalar field in the spatially regular curved spacetime
(\ref{Eq3}).

Our compact theorem therefore reveals the interesting fact that, as
opposed to stationary scalar fields
\cite{Fra,Lie,Notesr,Hern1,rfn1,rfn2,rfn3}, static self-gravitating
scalar fields whose nonlinear self-interaction potential $V(\psi^2)$
is a monotonically increasing function of its argument cannot form
spatially regular asymptotically flat boson stars.

\newpage

\bigskip
\noindent
{\bf ACKNOWLEDGMENTS}
\bigskip

This research is supported by the Carmel Science Foundation. I would
like to thank Yael Oren, Arbel M. Ongo, Ayelet B. Lata, and Alona B.
Tea for stimulating discussions.



\begin{thebibliography}{99}

\bibitem{Cha} J. E. Chase, Commun. Math. Phys. {\bf 19}, 276 (1970).

\bibitem{Bek1} J. D. Bekenstein, Phys. Rev. Lett. {\bf 28}, 452 (1972).

\bibitem{Teit} C. Teitelboim, Lett. Nuov. Cim. {\bf 3}, 326 (1972).

\bibitem{Heu} M. Heusler, J. Math. Phys. {\bf 33}, 3497 (1992);
M. Heusler, Class. Quant. Grav. {\bf 12}, 779 (1995).

\bibitem{Bek2} J. D. Bekenstein, Phys. Rev. D {\bf 51}, R6608 (1995).

\bibitem{BekMay} A. E. Mayo and J. D. Bekenstein, Phys. Rev. D {\bf 54}, 5059 (1996).

\bibitem{Bek20} J. D. Bekenstein, arXiv:gr-qc/9605059 .

\bibitem{Hodnmc} S. Hod, Phys. Lett. B {\bf 771}, 521 (2017).

\bibitem{NoteV} It is worth mentioning that, as noted in \cite{Bek20}, the characteristic
no-scalar-hair property of spherically symmetric static black holes
with regular event horizons is valid in the regime of static scalar
fields whose nonlinear self-interaction potential $V(\psi^2)$ is a
monotonically increasing function of its argument [see Eqs.
(\ref{Eq1}) and (\ref{Eq2}) below].

\bibitem{Hodrc} S. Hod, Phys. Rev. D {\bf 86}, 104026 (2012) [arXiv:1211.3202];
S. Hod, The Euro. Phys. Journal C {\bf 73}, 2378 (2013)
[arXiv:1311.5298]; S. Hod, Phys. Rev. D {\bf 90}, 024051 (2014)
[arXiv:1406.1179]; S. Hod, Phys. Lett. B {\bf 739}, 196 (2014)
[arXiv:1411.2609]; S. Hod, Class. and Quant. Grav. {\bf 32}, 134002
(2015) [arXiv:1607.00003]; S. Hod, Phys. Lett. B {\bf 751}, 177
(2015); S. Hod, Class. and Quant. Grav. {\bf 33}, 114001 (2016); S.
Hod, Phys. Lett. B {\bf 758}, 181 (2016) [arXiv:1606.02306]; S. Hod
and O. Hod, Phys. Rev. D {\bf 81}, 061502 Rapid communication (2010)
[arXiv:0910.0734]; S. Hod, Phys. Lett. B {\bf 708}, 320 (2012)
[arXiv:1205.1872]; S. Hod, Jour. of High Energy Phys. {\bf 01}, 030
(2017) [arXiv:1612.00014].

\bibitem{Herkr} C. A. R. Herdeiro and E. Radu, Phys. Rev. Lett. {\bf 112}, 221101
(2014); C. L. Benone, L. C. B. Crispino, C. Herdeiro, and E. Radu,
Phys. Rev. D {\bf 90}, 104024 (2014); C. A. R. Herdeiro and E. Radu,
Phys. Rev. D {\bf 89}, 124018 (2014); C. A. R. Herdeiro and E. Radu,
Int. J. Mod. Phys. D {\bf 23}, 1442014 (2014); Y. Brihaye, C.
Herdeiro, and E. Radu, Phys. Lett. B {\bf 739}, 1 (2014); J. C.
Degollado and C. A. R. Herdeiro, Phys. Rev. D {\bf 90}, 065019
(2014); C. Herdeiro, E. Radu, and H. R\'unarsson, Phys. Lett. B {\bf
739}, 302 (2014); C. Herdeiro and E. Radu, Class. Quantum Grav. {\bf
32} 144001 (2015); C. A. R. Herdeiro and E. Radu, Int. J. Mod. Phys.
D {\bf 24}, 1542014 (2015); C. A. R. Herdeiro and E. Radu, Int. J.
Mod. Phys. D {\bf 24}, 1544022 (2015); P. V. P. Cunha, C. A. R.
Herdeiro, E. Radu, and H. F. R\'unarsson, Phys. Rev. Lett. {\bf
115}, 211102 (2015); B. Kleihaus, J. Kunz, and S. Yazadjiev, Phys.
Lett. B {\bf 744}, 406 (2015); C. A. R. Herdeiro, E. Radu, and H. F.
R\'unarsson, Phys. Rev. D {\bf 92}, 084059 (2015); C. Herdeiro, J.
Kunz, E. Radu, and B. Subagyo, Phys. Lett. B {\bf 748}, 30 (2015);
C. A. R. Herdeiro, E. Radu, and H. F. R\'unarsson, Class. Quant.
Grav. {\bf 33}, 154001 (2016); C. A. R. Herdeiro, E. Radu, and H. F.
R\'unarsson, Int. J. Mod. Phys. D {\bf 25}, 1641014 (2016); Y.
Brihaye, C. Herdeiro, and E. Radu, Phys. Lett. B {\bf 760}, 279
(2016); Y. Ni, M. Zhou, A. C. Avendano, C. Bambi, C. A. R. Herdeiro,
and E. Radu, JCAP {\bf 1607}, 049 (2016); M. Wang, arXiv:1606.00811
.

\bibitem{Fra} F. E. Schunck and E. W. Mielke, Class. Quant. Grav. {\bf 20}, R301
(2003).

\bibitem{Lie} S. L. Liebling and C. Palenzuela, Living Rev. Relativity {\bf 15}, 6
(2012).

\bibitem{Der} G. H. Derrick, J. Math. Phys. {\bf 5}, 1252 (1964).

\bibitem{NoteTej} It is worth mentioning that Diez-Tejedor and
Gonzalez-Morales \cite{Tej} have recently provided an important
no-go theorem which explicitly reveals the fact that spatially
regular self-gravitating static scalar fields cannot form
spherically symmetric horizonless matter configurations which are
characterized by the metric property $g^{-1}_{00}(dg_{00}/dr)\geq0$.
In the present paper we shall provide a (remarkably compact) no-go
theorem which excludes the existence of horizonless boson stars made
of {\it static} scalar fields which is valid for generic values of
the physical quantity $g^{-1}_{00}(dg_{00}/dr)$.

\bibitem{Tej} A. Diez-Tejedor and A. X. Gonzalez-Morales, Phys. Rev. D {\bf 88}, 067302
(2013).

\bibitem{Noteunit} We use gravitational units in which $G=c=1$.

\bibitem{NoteSEH} Here $S_{\text{EH}}$ is the Einstein-Hilbert action.

\bibitem{Notesch} Here we use the Schwarzschild spacetime coordinates $(t,r,\theta,\phi)$.

\bibitem{Hodt1} S. Hod, Phys. Rev. D {\bf 84}, 124030 (2011)
[arXiv:1112.3286]; S. Hod, Phys. Rev. D {\bf 84}, 104024 (2011)
[arXiv:1201.0068]; S. Hod, Phys Lett. B {\bf 727}, 345 (2013)
[arXiv:1701.06587].

\bibitem{Notetag} Here a prime ${'}$ denotes a spatial derivative with respect to the radial coordinate $r$.

\bibitem{Notenbo} The near-origin relations (\ref{Eq5}) can be deduced from
the Einstein field equations, $G^{\mu}_{\nu}=8\pi T^{\mu}_{\nu}$,
which for spherically symmetric static spacetimes are given by the
functional relations \cite{BekMay,Hodt1} $\lambda'=8\pi
re^{\lambda}\rho-(e^{\lambda}-1)/r$ and $\nu'=(e^{\lambda}-1)/r+8\pi
re^{\lambda}p$, where $(\rho,p)\equiv (-T^{t}_{t},T^{r}_{r})$ are
the energy density and the radial pressure of the matter fields,
respectively. As explicitly proved in \cite{BekMay}, physically
acceptable spacetimes are characterized by finite values of the
energy-momentum components. Thus, taking cognizance of the above
stated Einstein equations and the relation (\ref{Eq4}), one obtains
the characteristic near-origin behavior (\ref{Eq5}) of the metric
functions.

\bibitem{Noteeg} Note that $e^{-\lambda}\neq0$ for spatially regular
horizonless spacetimes \cite{BekMay}.

\bibitem{Hodasm} S. Hod, Phys. Lett. B {\bf 739}, 383 (2014) [arXiv:1412.3808].

\bibitem{Noteano} Here we assume that $a\neq0$. If $a\equiv\psi(r\to0)=0$
then the asymptotic functional relation (\ref{Eq12}) informs us, in
accord with our assertion, that $\psi(r)$ is a non-monotonic
function of the radial coordinate $r$.

\bibitem{Notevdt} If $\dot V[\psi^2(r\to0)]=0$ then Eq. (\ref{Eq13})
yields the near-origin functional behavior $\psi(r\to0)=a/r$, which
violates the characteristic relation (\ref{Eq10}) of physically
acceptable systems.

\bibitem{Notesr} It is worth emphasizing the fact that
four-dimensional field solitons which violate the assumptions made
in the present theorem do exist as spatially regular mathematical
solutions of the non-linearly coupled Einstein-scalar field
equations. In particular, the composed Einstein-scalar system is
characterized by the presence of stationary (rather than static)
spatially regular solutions \cite{Fra,Lie,Hern1}, as well as by the
presence of non-asymptotically flat field configurations
\cite{rfn1,rfn2}. In addition, solitonic field configurations exist
for the case of scalar self-interaction potentials which are not
positive semidefinite \cite{rfn3}.

\bibitem{Hern1} C. A. R. Herdeiro, A. M. Pombo, and E. Radu, Phys.
Lett. B {\bf 773}, 654 (2017).

\bibitem{rfn1} Y. Brihaye, B. Hartmann, and S. Tojiev, Phys. Rev. D {\bf 88}, 104006
(2013).

\bibitem{rfn2} S. Ponglertsakul and E. Winstanley, Phys.
Lett. B {\bf 764}, 87 (2017).

\bibitem{rfn3} U. Nucamendi and M. Salgado, Phys. Rev. D {\bf 68}, 044026 (2003).

\end{thebibliography}
\end{document}